\documentclass[11pt,english]{article}
\pdfoutput=1
\usepackage{jheppub}
\usepackage{lscape}
\usepackage{amsmath}
\usepackage{slashed}
\usepackage{mathtools}
\usepackage{tikz-cd}
\usetikzlibrary{shapes,arrows,shadows}
\usepackage{amsmath,bm,times}
\usepackage{hyperref}
\usepackage{cleveref}
\usepackage{tcolorbox}
\makeatletter
\newcommand{\@fpheader}{}
\makeatother

\begin{document}

\title{Constructing allowed complex metrics from black holes}

\author[]{Oscar Loaiza-Brito, } 
\author[]{J. L. L\'opez-Pic\'on  and }
\author[]{Octavio Obregón}

\affiliation[]{Departamento de F\'isica, Universidad de Guanajuato, Loma del Bosque No. 103 Col. Lomas del Campestre C.P 37150 Leon, Guanajuato, Mexico.}

\emailAdd{oloaiza@fisica.ugto.mx, octavio@fisica.ugto.mx, jl\_lopez@fisica.ugto.mx}

\date{\today}
\today

\abstract
{We use diffeomorphic mappings to connect black hole metrics with complex solutions allowed by the Kontsevich-Segal criterion. By swapping radial and time-like coordinates and applying complex mappings, we derive dynamic metrics suitable for a Quantum Field Theory. This is shown for static and rotating black holes, mapping their interiors into the Kantowski-Sachs and a specific Gowdy-type cosmological model. We offer interpretations of the period during which the Kontsevich-Segal criterion holds.}


\maketitle{}

\section{Introduction}

The generation of new solutions to Einstein's equations through coordinate mappings on preexisting solutions has established itself as a particularly effective and fruitful approach in recent years. In the context of non-static models within the framework of General Relativity (GR), Kantowski-Sachs models represent an exceptional category within the spatially homogeneous Bianchi cosmological models. These models, notable for their intrinsic interest, have been investigated both as vacuum solutions and as fluid matter models, with and without a cosmological constant \cite{kantowski1966some}. The diffeomorphic transformation $t\leftrightarrow r$ that converts the Schwarzschild metric into the cosmological Kantowski-Sachs model is employed as a method to derive new cosmological solutions from those associated with black holes. This conceptual approach served as the foundation in \cite{obregon2001kerr},  resulting in a novel solution to Einstein's equations. This transformation facilitated the reinterpretation of the Kerr metric into what was, at the time, an unrecognized cosmological model within the Gowdy family. Such diffeomorphic alterations have also been independently utilized within string theory to correlate stationary solutions with time-dependent solutions \cite{leblond2003sd,garcia2005tachyon}.\\


In addition to the interchange between space and time coordinates, mappings to complex coordinates, such as the Wick rotation, have proven to be important in transforming Lorentzian classical metrics into  Euclidean solutions pertinent to path integrals designed to establish a consistent quantum field theory within curved backgrounds. In this regard, it is well-established that (semi)Euclidean metrics are employed to deduce significant physical results, such as the thermodynamic properties of black holes \cite{gibbons1977action,banados1992black}, whose Gaussian path integral exhibits convergence. The formal foundation of this transformation lies in statistical mechanics, specifically in the relationship between temperature and imaginary time \cite{wipf2013path}. Ultimately, this transformation functions as a mathematical technique with important implications in physics. In addition to the Wick rotation, similar complex transformations have been executed with the objective of yielding noteworthy physical results \cite{newman1965note,otherwickrotations2,otherwickrotations1,wen2024exactly,garnier2025complex,feldbrugge2017lorentzian}.\\

A more formal examination of the conditions for regularizing metrics within the framework of curved topological space-times was studied  in \cite{sorkin1997forks,louko1997complex}, and later pursued by Kontsevich and Segal \cite{kontsevich2021wick}, where a sufficient criterion was deduced for considering a complex metric permissible in the context of defining a consistent QFT in curved space. This condition shall hereafter be referred to as the Kontsevich-Segal (KS) criterion.\\

Although there is no proof that this criterion is necessary, interesting conclusions related to General Relativity, extending beyond cosmological solutions, are explored in \cite{briscese2022note,witten2022note, hertog2023kontsevich,hertog2025kontsevich,janssen2024ksw, Barbon:2025vvh}. Specifically, it was shown in \cite{witten2022note} that those metrics obtained from some complex transformations, which seem to be physically relevant,  indeed fulfill the KS criterion.  \\


In this context, for a quantum field theory (QFT)  to be {\it sufficiently} well-defined in  curved spacetime, it is desirable that the metric  satisfy two crucial conditions: it must be a solution to Einstein's equations and meet the KS criterion. Initially, satisfying both conditions presents a significant challenge. Nevertheless, it is feasible to construct permissible KS metrics from established solutions of Einstein's equations, particularly those involving black holes.\\

This work will adopt such an approach. Inspired by the ideas presented in \cite{obregon2001kerr}, where time-radius diffeomorphism in black hole metrics is utilized as a technique to derive new dynamical solutions to Einstein's equations, we will undertake further analytical continuations in non-temporal coordinates of these metrics. This will facilitate the generation of new, complex solutions that consequently comply with both Einstein's equations and the KS criterion.\\

Our work is organized as follows. In Section 2, we provide a review of the KS criterion and offer an  examination of static black hole solutions, demonstrating that the interior of a black hole can be utilized to derive a new metric satisfying the KS criterion. Section 3 focuses on deriving dynamical metrics from static black hole solutions through the introduction of the well-established diffeomorphic mapping between time and radius. Subsequently, an analogous mapping is employed to render these dynamical metrics admissible within the corresponding horizons. Notably, we observe a correlation between the duration in which the solutions meet the KS criterion and the classical break-time as proposed by Dvali \cite{Dvali:2017eba}. In Section 4, the same mappings are applied to the Kerr space-time. The dynamical metric associated with this black hole was previously linked to a cosmology of the Gowdy type, and the proposed mappings render this metric permissible in a critical region of space-time inside the horizon. Section 5 is dedicated to discussions and conclusions.







\section{Kontsevich-Segal Criterion and allowable static metrics}
Let us briefly summarize the Kontsevich-Segal (KS) Criterion.  For that, consider a free  gauge field $F_{\mu_1\dots\mu_q}=\partial_{[\mu_1}A_{\mu_2\dots\mu_q]}$ on a $4$-dimensional space-time, with an action 
\begin{align}
    I_q=\kappa_q\int F_{\mu_1\dots\mu_q}F^{\mu_1\dots\mu_q}\, d^qx,
\end{align}
for $0\leq q\leq 3$ and $\kappa_q$ the corresponding coupling constant. It was shown in \cite{kontsevich2021wick} that the path integral of this field would be convergent if 
\begin{align}
    \text{Re}( F_{\mu_1\dots\mu_q}F^{\mu_1\dots\mu_q})>0.
    \label{C1}
\end{align}
This is extended to $q=0$ by considering  that, in such a case, we have a well-defined scalar field theory at the quantum level. On the other hand, the convergence of the path integral leads to a well-defined quantum theory in curved space-time. However, in general, condition (\ref{C1}) depends on a complex metric,  implying that the criterion can be translated in terms of constraints on it. Indeed, it was proved in \cite{kontsevich2021wick} that a complex metric is allowable for use in the corresponding path integral if
\begin{align}
    \sum_{\mu=0}^{3}|\text{Arg}~\lambda_\mu |<\pi,
\end{align}\\
where $\lambda_\mu$ are the eigenvalues of the complex metric $\mathfrak{g}$. Henceforth, we shall refer to this constraint as the Kontsevich-Segal (KS) criterion.\\

Notice that any complex metric fulfilling the KS criterion could represent a curved background where a QFT is well defined. Therefore, it is of great importance to develop methods to construct allowable metrics, including the construction of dynamical ones. In the following sections, we present a method based on the use of internal black hole solutions to generate KS allowable dynamical backgrounds.\\

\subsection{Static metric from Schawrzschild black hole}
As shown in \cite{witten2022note}, complex solutions of Einstein's equations that could lead to potential physical problems seem to violate the KS criterion. On the contrary, it is argued that Black hole solutions which undergo  a Wick rotation satisfying the KS criterion allow us to obtain the thermodynamic properties of black holes from the path integral.\\

Hence, our approach is to use black hole solutions as a reliable source to generate complex metrics that fulfill the KS. For that, it is interesting to consider a spherically symmetric metric with time and radial components depending on $r$ 
\begin{equation}
    ds^2=-e^{2\lambda(r)}dt^2+e^{2\nu(r)}dr^2+r^2d\theta^2+r^2\sin^2\theta d\phi^2,
\end{equation}
such that it possesses a horizon; this is $r>r_\ast$, for some $r_\ast$, 
\begin{align}
   e^{2\lambda(r)}>0 \qquad\text{and}\qquad e^{2\nu(r)}>0
\end{align} and the contrary for $r<r_\ast$. 
Since the  metric is Lorentzian for  $r>r_\ast$, after a Wick rotation $t\rightarrow i\tau$, the metric becomes
\begin{align}
    ds^2=e^{2\lambda(r)}d\tau^2+e^{2\nu(r)}dr^2+r^2d\theta^2+r^2\sin^2\theta d\phi^2.
\end{align}
The metric components are all real and positive; therefore, $\text{Arg}(g_{\mu\mu}) = 0$ for all $\mu$, and Eq. (1) is satisfied. However, inside the horizon, this is for $r<r_\ast$,  the KS criterion cannot be fulfilled since  $g_{tt}$ and $g_{rr}$ are negative. The prototypical example is, of course, the Schwarzschild metric.\\


One can think of the construction of an allowable complex metric around a Schwarzschild black hole. Outside the horizon, it seems plausible to have the necessary conditions to construct such a metric \cite{witten2022note}. Let us briefly review the analytical continuation of the Schwarzschild metric outside the horizon and  how its Euclidean version allows us to derive some thermodynamical properties. \\

Taking $\lambda(r)=1/\nu(r)=\ln(1-2M)$,
we  obtain the Schwarzschild solution of a static black hole, which, upon Wick's rotation, becomes

\begin{align}
ds^2 = \left( 1-\frac{2M}{r} \right)d\tau^2 + \frac{1}{\left( 1-\frac{2M}{r} \right)}dr^2 + r^2d\theta^2 + r^2 \sin^2{\theta}\, d\phi^2.
\end{align} 
Close to the horizon $r_\ast=2M$, the metric has a conical singularity that is resolved if $\tau$ is periodic; this is
\begin{align}
    \tau\sim \tau + 8\pi M,
\end{align}
where the period is inverse to the Hawking temperature $T=1/8\pi M$, allowing the Euclidean Schwarzschild solution to be a smooth manifold without singularities at $r=2M$ and ensuring consistency in the Euclidean path integral formulation. The horizon acts as a saddle point for the complex metric, fulfilling the KS criterion since the metric components are real for $r>2M$. The extension of the KS criterion to semi-classical gravitational systems, discussed in \cite{witten2022note}, connects it to saddle points like black hole horizons, offering thermodynamical insights upon euclideanization.\cite{Gibbons:1976ue}. \\

How should one interpret this extension of the KS criterion to quantum gravity? Basically, as stated in \cite{witten2022note}, at least for the case of the Schwarzschild Black hole, this means that the time component of the metric is always negative outside the horizon, allowing for the possibility of consistently defining a time-like vector. The same question can be addressed for other types of black holes, as we shall discuss in the next section.\\

\subsection{Constructing allowed static metrics from a black hole interior}
We want to focus on the possibility of using the interior of a black hole as a source for generating a metric that fulfills the KS criterion. For that purpose, we take the Schwarzschild metric and perform the analytical continuation through the  mappings $t\rightarrow i\tau$ and $\theta\rightarrow i\delta$. The final metric, after multiplying by a factor of -1 (which does not alter the fact that it is a solution of Einstein's equations), reads
\begin{equation}
    ds^2= \left(\frac{2M}{r}-1 \right)d\tau^2 + \frac{1}{\left(\frac{2M}{r}-1 \right)}dr^2 + r^2d\delta^2 + r^2 \sinh^2{\delta} ~d\phi^2.
\end{equation}

This metric satisfies the KS criterion for $0<r<2M$. Another way to see this is to realize that close to $r=2M$, this metric also has a conical singularity and is equally resolved for $\tau$, being periodic. However, the singularity at $r=0$ is not resolved. \\

Hence, it seems at least that for static, radially dependent, angularly symmetric metrics, where there is a horizon defined, the KS criterion is satisfied by two different metrics defined at different values of $r$, separated by the horizon or the saddle point in the complex analytical continuation that connects both. Notice that we are not providing an allowed KS metric inside the black hole; we are merely constructing a metric that fulfills the KS criterion for $r<2M$.\\

One can also consider a metric with hyperbolic symmetry instead of spherical, in the form

\begin{equation}
    ds^2=-e^{2\lambda(r)}dt^2+e^{-2\lambda(r)}dr^2+r^2d\theta^2+r^2\sinh^2\theta d\phi^2.
\end{equation}
By applying the same mappings, we end up with an allowed KS metric, characterized by spherical symmetry, given by
\begin{equation}
     ds^2= e^{2\lambda(r)}d\tau^2 + e^{-2\lambda(r)}dr^2 + r^2d\delta^2 + r^2 \sin^2{\delta} \,d\phi^2,
\end{equation}
with $\lambda(r)>0$.

\subsection{de Sitter space}
Let us consider another static metric with a horizon. In this case, we consider the de Sitter (dS) metric in static coordinates given by
\begin{align}
    ds^2= -\left( 1-\frac{r^2}{3\Lambda} \right)dt^2+\frac{1}{(1-\frac{r^2}{3\Lambda})}dr^2+r^2d\Omega^2.
\end{align}
In this case, after a Wick rotation, the KS criterion is fulfilled for $r<3\Lambda$, indicating that it is possible to define a QFT over the Euclidean version of this space-time for any $r$ inside the horizon. This represents the idea that the entire visible universe could be an allowable background for constructing a quantum field theory at a given time. Near the horizon, we can also find that the complex time $\tau$ behaves as an angular component of a smooth two-dimensional manifold, whose singularity is resolved if $\tau$ is periodic with period
\begin{align}
    \tau\rightarrow \tau + 24\pi\Lambda.
\end{align}
In this sense, we can assign a Hawking temperature at the horizon $T=1/24\pi\Lambda$. The smaller the cosmological constant, the highest the Hawking temperature.

    \section{Allowable Dynamical metrics}
We have used the interior of a black hole to construct allowable KS complex static metrics by implementing some analytical continuation or complex mappings on certain coordinates. However, we can also use the interior metric of a black hole to construct a dynamical metric based on the interchange between radial and temporal coordinates, as was done to relate the interior of a Schwarzschild black hole to the Kantowski-Sachs cosmological solution, or by relating the interior of a Kerr black hole to a new type of cosmological Gowdy-like solution, as done in \cite{obregon2001kerr}. This map leads us to a Lorentzian dynamical metric, which, for some scenarios, can be identified with a cosmological background. It is, then, interesting to think about the physical meaning of a dynamical metric that fulfills the KS criterion.  We shall discuss this, starting with some well-known dynamical metrics, such as the dS and the FRW metrics. \\

\subsection{Dynamical de Sitter}
The de Sitter space in static coordinates can be transformed into a time-dependent one, describing the space-time seen from the perspective of a freely falling observer. In general, the associated metric involves the FRW metric, which, after being analytically continued by $t=\rightarrow i\tau$, reads
\[
ds^2 = d\tau^2 + a^2(i\tau) \left[ \frac{dr^2}{1 - kr^2} + r^2 \left(d\theta^2 + \sin^2\theta \, d\phi^2 \right) \right].
\]

Consider the case for $k=0$. Since the metric is diagonal, it is easy to see that the sum of the arguments  is $6~|\text{Arg}(a(i\tau)|$. Notice that, for a matter- or radiation dominated universe, the scale factor $a(t)$ goes as $t^{2/3}$ and $t^{1/2}$ respectively, for which the KS criterion is not fulfilled. However, for an exponentially accelerated universe, or dS, $a(i\tau)=e^{iH\tau}$, the KS criterion is satisfied for 
\begin{align}
\tau<\frac{\pi}{6H}.
\end{align}

After all, this seems to be consistent. The KS criterion, which is a necessary condition  to define a QFT on a complex metric, is a mathematical constraint upon which a classical Einstein solution can still be adequate for defining the quantum dynamics of a field. However, in the case of a dynamical metric, where the condition is translated into a bound on time, we can interpret it as the time at which the quantum dynamics of the metric itself, viewed as a system formed by quantum states (perhaps a coherent state), departs from the evolution of the classical mean fields. In other words, the classical description of the metric background is no longer compatible with a quantum picture. This resembles the classical-break time for de Sitter space, as proposed in \cite{Dvali:2017eba}, where the dS space is seen as an excited coherent state of gravitons over Minkowski space. A similar scale of time was obtained in \cite{Damian:2024vmv} by considering a coherent state of states within the context of string theory.\\



\subsection{From Schwarzschild to Kantowski-Sachs}
Let us construct dynamical metrics departing from the interior solution of a BH. We are interested in finding allowable cosmological metrics, particularly those metrics obtained by interchanging the time and radial coordinates in the interior black hole solution. The cosmological Kantovski-Sachs metric is obtained by mapping variables $r \rightarrow t$ in the Schwarzschild metric for $r<2M$, from which 

\begin{align}
ds^2 =  -\frac{1}{\left(\frac{2M}{t} -1\right)}dt^2 + \left(\frac{2M}{t} -1 \right)dr^2 + t^2d\theta^2 + t^2 \sin^2{\theta} d\phi^2,
\label{KS}
\end{align} 

\noindent where we now have a non-static cosmological metric for $0 < t < 2M$. It can be shown that this metric satisfies Einstein's equations. Hence, the question we want to pose is whether there exist mappings that transform this metric into one fulfilling the KS criterion.\\ 

There are different ways we can construct a complex KS allowed metric from (\ref{KS}). We present two significant mappings \cite{otherwickrotations1}:

\begin{enumerate}
    \item Take  $~r \rightarrow i \rho, ~\theta \rightarrow i \delta$ and $~g_{\mu \nu} \rightarrow  - g_{\mu \nu}$. The resulting metric is  a solution of Einstein's equations given by
    \begin{align}
ds^2 =  \frac{1}{\left(\frac{2M}{t} -1\right)}dt^2 + \left( \frac{2M}{t} -1 \right)d \rho^2 + t^2d\delta^2 + t^2 \sinh^2{\delta} d\phi^2,
\end{align}

 	
	

	



These mappings change the topology of the angular metric from spherical to hyperbolic. It can be noticed that  in the region $0 < t < 2M$ the metric is 
Euclidean, and hence the Kontsevich-Segal criterion is trivially satisfied.  

\item Consider the maps $t\rightarrow i\tau, ~r \rightarrow i \rho$ and $~\theta \rightarrow i \delta$. The resulting metric is given by
\begin{align}
    ds^2 = \left(1 - \frac{2M}{i\tau}\right)d\rho^2 - \left(1 - \frac{2M}{i\tau}\right)^{-1}d\tau^2 + \tau^2 d\delta^2 + \tau^2 \sinh^2\delta \, d\phi^2,
\end{align}
from which

\begin{align}
\sum_{\mu=0}^3 |\text{Arg}(\lambda_{\mu})|=2 \tan^{-1}\left(\frac{2M}{\tau}\right),
\end{align}

which is always less than $\pi$ for all  $\tau> 0$. Then, this is a metric that is a solution to the Einstein equation, is time-dependent, and fulfills the KS criterion. Therefore, it represents a dynamical background on which it is possible to construct a well-defined quantum field theory for all times.\\

\item Take the transformation $r\rightarrow i\rho$. The resulting metric reads
\begin{align}
ds^2 =  -\frac{1}{\left(\frac{2M}{t} -1\right)}dt^2 - \left(\frac{2M}{t}-1 \right)d\rho^2 + t^2d\theta^2 + t^2 \sin^2{\theta} d\phi^2.
\end{align} 
Since this metric is diagonal, by explicitly applying the criteria in Eq. (1), we have that

\begin{equation}
\sum_\mu |\text{Arg}(\lambda_{\mu})| = \bigg \{\
\begin{array}{ccc}
2\pi ~~~~~~~ 0 < t < 2M \\
0 ~~~~~~~~~~~~ t > 2M.
\end{array}
\end{equation}
This background represents an allowable metric $t >2M$. It is a dynamical metric on which the quantum dynamics cannot be properly defined at earlier times. 
\end{enumerate}

\subsection{Allowable Dynamical metric from Kerr}
The rotating black hole, described by the asymptotically flat Kerr metric, fails to meet the KS criterion within the ergospheres (after the application of Wick rotation). This result is briefly reviewed in Appendix A. Instead, a suitable curved background for quantum field theory is provided by the asymptotically AdS Kerr black hole, where the KS criterion is satisfied outside the event horizon \cite{witten2022note}. \\

Nonetheless, let us take advantage of this scenario, analogous to the approach used for the Schwarzschild Black Hole. Using the metric within the interior of the Kerr Black Hole, it is possible to construct an alternative metric which, upon appropriate analytic continuation, results in a complex metric satisfying the KS criterion in a sensitive region inside Kerr Cosmology. To achieve this, the Kerr metric in its original coordinates $(t, r, \theta, \phi)$ is considered, focusing on the range for $r$ situated between the ergospheres, where $g_{tt}>0$.\\

Hence, the Kerr metric in Boyer-Lindquist coordinates is given by
\begin{equation}
ds^2= -\frac{\Delta}{\Sigma}(dt-a \sin^2\theta d\phi)^2+\frac{\sin^2\theta}{\Sigma}\left((r^2+a^2)d\phi-a dt\right)^2+\frac{\Sigma}{\Delta}dr^2+\Sigma d\theta^2,
\label{Kerr-BL}
\end{equation}
where $\Sigma$ and $\Delta$ are
\begin{equation}
\Sigma(r)=r^2+a^2\cos^2\theta, \qquad \Delta(r)= r^2-2Mr+a^2,
\end{equation}
with $a$ the angular momentum.\\

Now,  let us turn to the dynamical metric by exchanging the time variable $t\leftrightarrow r$. The resulting metric reads

\begin{align}
ds^2 = & - \left( \frac{t^2 + a^2\cos^2{\theta}}{2Mt-t^2-a^2} \right)dt^2 + \sin^2{\theta}\left[ t^2 + a^2 + \frac{2Ma^2 t \sin^2{\theta}}{t^2 + a^2\cos^2{\theta}} \right] d\phi^2 - \left( \frac{2Mat\sin^2{\theta}}{t^2 + a^2\cos^2{\theta}} \right)drd \phi \\ \nonumber
& + \left( \frac{2Mt}{t^2 + a^2\cos^2{\theta}} -1 \right) dr^2 + (t^2 + a^2\cos^2{\theta})d\theta^2.
\label{Kerr2}
\end{align} 

In contrast to the Schwarzschild metric, this metric exhibits a non-diagonal form and has been identified in \cite{obregon2001kerr} as a cosmological metric within the Gowdy family.
Consider now the complex mappings $r \rightarrow iM\rho$, $\theta \rightarrow i\delta$, and $g_{ij} \rightarrow -g_{ij}$ for which the metric \ref{Kerr2} becomes

\begin{align}
ds^2 = & \left( \frac{T^2 + \lambda^2\cosh^2{\delta}}{2T-T^2-\lambda^2} \right)M^2dT^2 + \left(\frac{2T}{T^2 + \lambda^2\cosh^2{\delta}} -1 \right) M^2 d\rho^2  - \left( \frac{2i \lambda T\sinh^2{\delta}}{T^2 + \lambda^2\cosh^2{\delta}} \right)M^2 d\rho d \phi \\ \nonumber 
& + (T^2 + \lambda^2\cosh^2{\delta})M^2d\delta^2 + \sinh^2{\delta}\left[ T^2 + \lambda^2 - \frac{2\lambda^2 T \sinh^2{\delta}}{(T^2 + \lambda^2\cosh^2{\delta})} \right]M^2 d\phi^2,
\label{finalm}
\end{align} 
where $\lambda=a/M$ and $T=t/M$. The eigenvalues of this metric are

\begin{align}
& \lambda_0 = G_{tt} = M^2\left( \frac{T^2 +\lambda^2\cosh^2{\delta}}{2T-T^2-\lambda^2} \right),~~~~\lambda_1 = G_{\delta \delta} = M^2(T^2 + \lambda^2\cosh^2{\delta}) \\ \nonumber
& \lambda_{2,3} = \frac{G_{\rho \rho} + G_{\phi \phi}}{2} \pm \frac{\sqrt{(G_{\rho \rho} + G_{\phi \phi})^2-4(G_{\rho \rho}G_{\phi \phi} - G_{\rho \phi}^2)}}{2},
\end{align}
where $G_{\mu\nu}$ are the corresponding metric components.\\

A notable distinction from the Kantowski-Sachs metric is that this dynamical metric possesses explicitly complex components, and the dependence on the angle $\delta$ is more intricate than in the Schwarzschild-Kantowski-Sachs case. In the general case, the KS criterion reads\footnote{It is important to note that the function $\tan^{-1}(x)$ depends on  the quadrant where the complex number is located. Therefore, if the real part of the complex number is negative, an additional angle of $\pi$ must be added to each eigenvalue in equation (4.13).}


\begin{align}
\tan\left(\frac{1}{2}\sum_{\mu=0}^3 |\text{Arg}(\lambda_{\mu})\right)
=\frac{{\sqrt{| 4(T^2+\lambda^2-2T)\sinh^2{\delta}+\left( \frac{2T-2\lambda^2T\sinh^4{\delta}}{T^2+\lambda^2\cosh^2{\delta}} + (\lambda^2+T^2)\sinh^2{\delta}-1 \right)^2}| }}{\frac{2T-T^2+T^2(\lambda^2+T^2)\sinh^2{\delta}-2\lambda^2T\sinh^4{\delta}
+\lambda^2[(\lambda^2+T^2)\sinh^2{\delta}-1]\cosh^2{\delta}}{T^2+\lambda^2\cosh^2{\delta}}}.
\end{align}

\begin{figure}[htbp]
  \centering
  \includegraphics[width=0.7\linewidth]{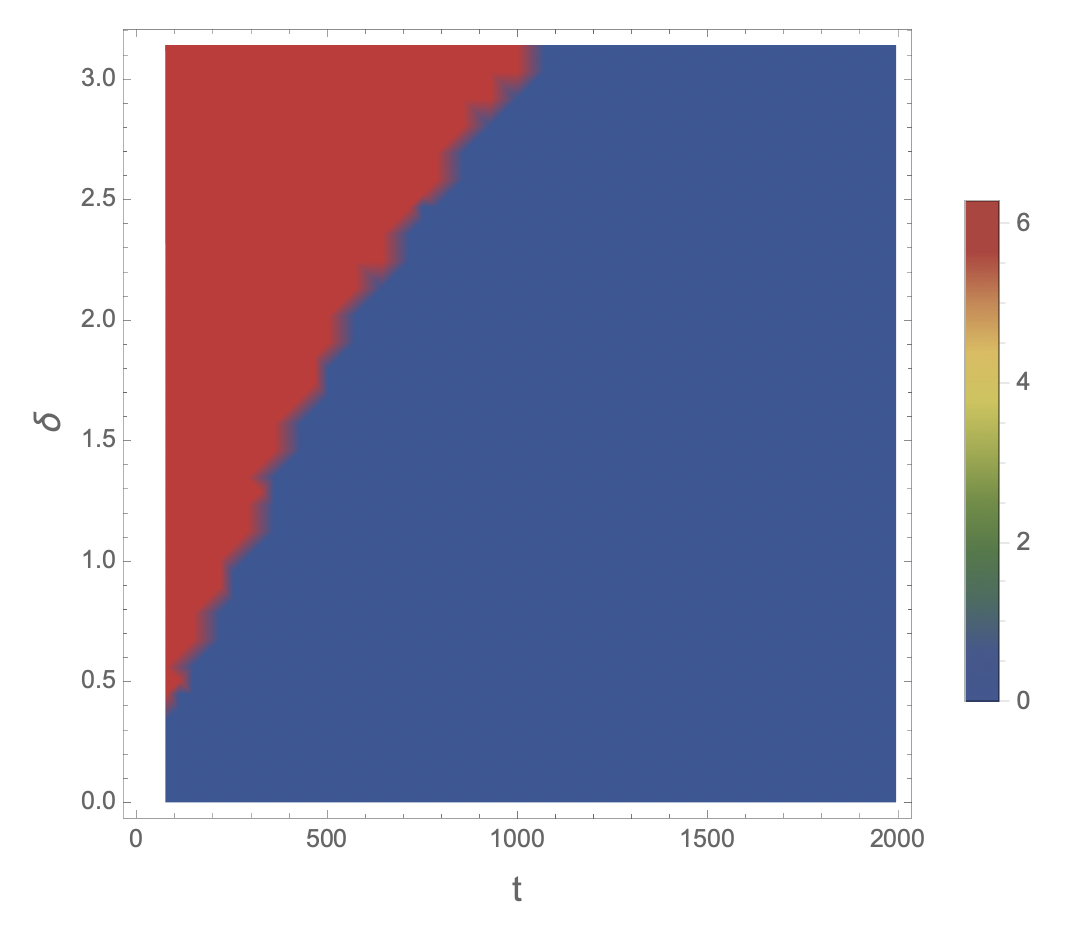}
  \caption{This figure shows a density plot of the function in Eq. (4.13) for the particular values $M=1000$, $a=100$ and ratio 
  $\lambda =0.1$. In this case the location of horizons in non-reduced units is; $t_{-} = 5.012562893$, $t_{+}=1994.9874371066$ and $t_0=1066.09$. In the region $t_0 < t < t_{+}$ the criterion is met for all values of the angle $\delta$.}
  \label{Fig1}
\end{figure}

The main conclusion is that, for fixed values of the angular momentum $a$ and the mass $M$, there exists an interval of $\delta$ for which the KS criterion is satisfied. This behavior is shown in Figure (\ref{Fig1}), where $M=1000$ and $a=100$ are kept fixed. One can see that the domain in which the criterion holds depends on the ratio of angular momentum to mass. As this ratio becomes smaller, the admissible region grows, so that the dynamical metric is well defined in a certain range between the time horizons, whose positions, labeled $T_{\pm}$, are given by
\begin{align}
    T_\pm=1\pm\sqrt{1-\lambda^2}.
\end{align}

 There exists a 
 time $t_0$ at which the region that lies between $t_{-}$ and $t_{0}$ ceases to be valid. $t_0$ approaches $t_{-}$ as the ratio $a/M$ decreases, namely $\lambda \rightarrow 0$. Hence, the valid region in the interval $t_{0} < t < t_{+}$ (in non-reduced units) increases as a function of $\lambda$. Let us define the percentage fraction of the valid region as

\begin{equation}
P(\lambda) = 1- \frac{t_{ 0} - t_{-}}{t_{+} - t_{-}}
\end{equation}

This percentage remains constant for the same ratio $\lambda$ and tends to 1 as the ratio decreases, namely in the limit $a\rightarrow 0$. 
This behavior is shown in figure 2. Figure 1 shows a density mapping of the function defined by equation $(3.13)$ for specific values of $M = 1000$ and $a = 100$, for a ratio $\lambda =0.1$. In this particular case, the horizons are located at $t_{-} = 5.013$ and $t_{+} = 1994.99$, and the value of $t_{0} = 1066.09$, for a total criterion validity region of approximately 47 percent. In Figure 1, a vertical line can be drawn starting from the value at t = 1066.09, and in the region to the right of this vertical line, the function defined in Equation 3.13 is strictly less than $\pi$, hence satisfying Eq. (2.3). \\

\begin{figure}[htbp]
  \centering
  \includegraphics[width=0.7\linewidth]{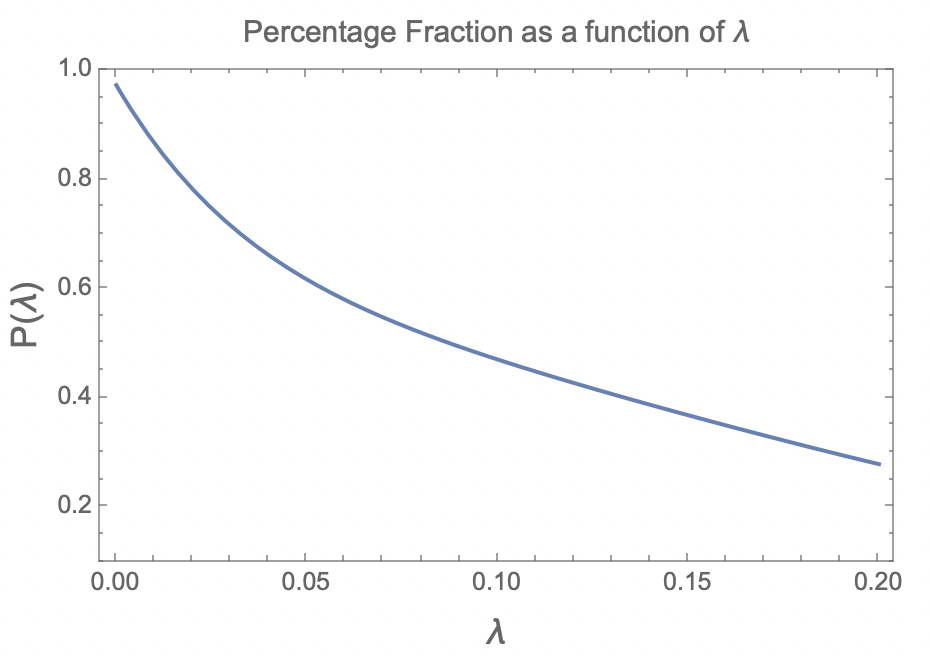}
  \caption{This figure shows the raise of the percentage of validity of the criterion within the horizons as the ratio $\lambda$ decreases, namely in the limit $a\rightarrow 0$. The particular value $t_0$ at which the criterion ceases to be valid approaches the singularity.}
  \label{Fig2}
\end{figure}

Finally, we would like to see if the metric (\ref{finalm}) corresponds to some Wick rotation of a Lorentzian metric. Indeed, a Lorentzian metric is given by

\begin{align}
ds^2 = & -\sinh^2{\delta}\left[ T^2 + \lambda^2 - \frac{2\lambda^2 T \sinh^2{\delta}}{(T^2 + \lambda^2\cosh^2{\delta})} \right]M^2d\varphi^2 + \left( \frac{2 \lambda T\sinh^2{\delta}}{T^2 + \lambda^2\cosh^2{\delta}} \right)M^2 d\rho d \varphi\nonumber\\ 
&+ \left(\frac{2T}{T^2 + \lambda^2\cosh^2{\delta}}-1 \right) M^2d\rho^2
+\left( \frac{T^2 + \lambda^2\cosh^2{\delta}}{2T-T^2-\lambda^2} \right)M^2dT^2+ (T^2 + \lambda^2\cosh^2{\delta})M^2d\delta^2,
\label{finalm2}
\end{align} 
 leads us to \ref{finalm} under the mapping $\varphi \rightarrow i \phi$. This last metric is a static one, with a time coordinate given by $\varphi\in [0, 2\pi)$.\\

 Assuming that a Lorentzian metric derived from black hole solutions via a Wick rotation, which complies with the KS criterion, is valid for elucidating the physical characteristics of such a metric, this static metric can be considered a permissible background framework for the formulation of Quantum Field Theory (QFT).

\section{Final comments and conclusions}

Through the exchange of radial and time-like coordinates, coupled with the application of complex mappings on Black Hole metrics, we construct dynamical metrics satisfying the Kontsevich-Segal criterion. This indicates that the resulting solutions could serve as well-defined, curved, and time-dependent metrics upon which a quantum field theory could be  formulated.\\

We provided some interesting dynamical cases. For example, by applying a Wick rotation to the FWR metric for an accelerated universe, we note that the KS criterion is fulfilled for a finite lapse of time, which is given as $\tau\sim 1/H$, where $H$ is the Hubble constant. This result resembles the classical break-time proposed in \cite{Dvali:2017eba}, representing the moment when the dynamical de Sitter classical solution departs from the evolution of a coherent state of gravitons, indicating that  the classical background does not describe the evolution of a quantum system.\\

Moreover, we construct dynamical complex metrics that satisfy the KS criterion using internal structures arising from Black Hole solutions.
Additionally, we focus on the Schwarzschild and Kerr Black Holes as primary examples.
Relying on the interchange between the radial and time coordinates, we incorporate specific complex mappings.
These transformations allow us to generate nontrivial complexified geometries that, in the case of the Kerr metric, can be produced by a Wick transformation of a static metric, in which the axial angular coordinate is now identified as the time coordinate.
Hence, the resulting metrics may serve as viable backgrounds for quantum field theories.
As such, they present a consistent framework for further theoretical exploration.\\

\begin{center}{\bf Acknowledgements}
\end{center}

We thank Nana Cabo-Bizet and Hugo Garc\'ia-Compe\'an for their useful suggestions and comments. O. Obregón and J. L. López-Picón were supported by SECIHTI Project: Ciencia Básica y de Frontera CB2023-2024-2923. \\

\appendix

\section{Kerr metric and the KS criterion}
As mentioned in \cite{witten2022note}, the pseudo euclidean version of the asymptotically flat Kerr metric does not fulfill the KS criterion. In this section, we review some details about this result.\\

For the metric (\ref{Kerr-BL}), the horizons are defined by $\Delta=0$ and are given by
\begin{equation}
r_\pm=M\pm\sqrt{M^2-a^2}.
\end{equation}

Following \cite{witten2022note}, we can express this metric  in the form of
\begin{equation}
ds^2= -N^2dt^2+\rho^2(N^\phi dt+d\phi)^2+g_{rr}dr^2+g_{\theta\theta}d\theta^2, 
\end{equation}
with
\begin{eqnarray}
g_{tt}&=&-N^2+\rho^2(N^\phi)^2=\frac{-\Delta+a^2\sin^2\theta}{\Sigma}=-1+\frac{2Mr}{\Sigma}\nonumber\\
g_{t\phi}&=&-\rho^2N^\phi=\frac{a\sin^2\theta}{\Sigma}\left(\Delta -(r^2+a^2)\right)=-\frac{2Mra~\sin^2\theta}{\Sigma}\nonumber\\
g_{\phi\phi}&=&\rho^2=\frac{\sin^2\theta}{\Sigma}\left(-\Delta a^2 \sin^2\theta+(r^2+a^2)^2\right)\nonumber\\
g_{rr}&=&\frac{\Sigma}{\Delta}\nonumber\\
g_{\theta\theta}&=&\Sigma,
\end{eqnarray}
from which
\begin{equation}
N^\phi= -\frac{g_{t\phi}}{g_{\phi\phi}}=\frac{2Mra}{\Delta\Sigma+2Mr(r^2+a^2)},
\end{equation}
with $\lim_{r\rightarrow \infty}N^\phi=0$. $N^\phi$ evaluated at the external horizon $r_+$ defines the angular velocity
\begin{equation}
\Omega_+=\frac{a}{r_+^2+a^2}.
\end{equation}

The ergospheres $r_{e\pm}$ are defined by $g_{tt}=0$ and are given by
\begin{eqnarray}
r_{e+}&=& M+\sqrt{M^2-a^2\cos^2\theta}\ge r_{+},\nonumber\\
r_{e-}&=& M-\sqrt{M^2-a^2\cos^2\theta}\le r_{-}.
\end{eqnarray}

First of all, notice that for $r_+<r<r_{e+}$, the metric component $g_{tt}$ is non-negative while it vanishes at the external ergosphere $r_{e+}$, and $g_{tt}<0$ for $r> r_{e_+}$. This implies that  an observer in the ergosphere cannot  stand still while having a time-like trajectory. Therefore, the region where an observer can stand still is defined by $r> r_{e+}$.\\

One could think that this is remedied by adding angular velocity to the observer, implying a change in coordinates.  We select a
coordinate transformation given by $\tilde\phi=\phi+\Omega_{+}t$ for which the metric reads
\begin{equation}
ds^2=\left(-N^2+\rho^2(N^\phi-\Omega_{+})^2\right)dt^2+2\rho^2(N^\phi-\Omega_{+})dtd\tilde\phi+ \rho^2d\tilde\phi^2+g_{rr}dr^2+g_{\theta\theta}d\theta^2.
\end{equation}
Notice that at the external horizon $r_{+}$, the component of the metric $g_{t\phi}$ vanishes; however, an observer can stand still only in regions where $g_{tt}<0$, which occurs when  $N^2>\rho^2(N^\phi-\Omega_+)^2$. In other words, notice that  $g_{tt}(r_+)=0$ and for very large $r$, 

\begin{equation}
g_{tt}\sim -1-2\Omega_+^2+(r^2-a^2)\Omega_+^2 \sim \Omega_+^2 r^2,
\end{equation}
indicating that at some point outside the external horizon, $g_{tt}$ becomes positive depending on the value of the angular momentum $a$.\\

Performing the analytical continuation  by the Weyl transformation $t\rightarrow i\tau$, we get
\begin{equation}
ds^2=\left(N^2-\rho^2(N^\phi-\Omega_{+})^2\right)d\tau^2+2i\rho^2(N^\phi-\Omega_{+})d\tau d\tilde\phi+ \rho^2d\tilde\phi^2 + \frac{\Sigma}{\Delta}dr^2+\Sigma d\theta^2.
\label{complexkerrm}
\end{equation}

The eigenvalues of this quasi-Euclidean metric are given by
\begin{eqnarray}
\lambda_{1,2}&=&\frac{1}{2} \left(\text{Tr}\mathcal{M}\pm\sqrt{(\text{Tr}\mathcal{M})^2-4\text{det}\mathcal{M}}\right)\nonumber\\
\lambda_3&=&\Sigma\nonumber\\
\lambda_{4}&=&\frac{\Sigma}{\Delta},
\end{eqnarray}
where
\begin{equation}
\mathcal{M}=
\begin{pmatrix}
N^2-\rho^2(N^\phi-\Omega_{+})^2\,\,&i\rho^2(N^\phi-\Omega_{+})\\
i\rho^2(N^\phi-\Omega_{+})&\rho^2
\end{pmatrix}
\end{equation}
from which
\begin{equation}
\text{det}(\mathcal{M})=\rho^2N^2.
\end{equation}
From these results, we obtain
\begin{eqnarray}
\text{Arg}(\lambda_{1,2})&=&\tan^{-1}\left(\pm\frac{\text{Im}(\sqrt{(\text{Tr}(\mathcal{M}))^2-4~\text{det}(\mathcal{M})}}{\text{Tr}(\mathcal{M})}\right)\nonumber\\
\text{Arg}(\lambda_3)&=&\tan^{-1}(\Sigma)=0,\nonumber\\
\text{Arg}(\lambda_4)&=&\tan^{-1}\left(\frac{\Sigma}{\Delta}\right)=0,
\end{eqnarray}
since $\Sigma$ and $\Delta$ are always positive (notice that $\Delta(r_{e+})=a^2)$. \\

This metric is then an allowed metric if and only if both eigenvalues $\lambda_{1,2}$ are real and positive, or if they form a conjugate pair with positive real parts. But in either scenario, this happens only if $\text{Tr}(\mathcal{M})$ and $\text{det}({\mathcal M})$ are both positive. Since for large $r$, ${\mathcal M}_{11}$ is negative, it could be that the trace is negative. Therefore, we conclude that the complex metric \ref{complexkerrm} is not allowed for some $r$ outside the horizon \cite{witten2022note}. Hence, we see that in both coordinate systems $(t, r, \theta, \phi)$ or $(t, r, \theta, \tilde\phi)$ there are regions for $r>r_+$ where $g_{tt}$ is positive, rendering the KS criterion unfulfilled.\\

\bibliographystyle{JHEP}
\bibliography{references2}  
\end{document}